\documentclass[preprint]{aastex}

\def\sunindex{\hbox{$_\odot$\ }}

\def\scnb{$\rm{S}(3839)$\ }
\def\scnr{$\rm{S}(4142)$\ }
\def\sch{$\rm{CH}(4300)$\ }

\shorttitle{CN variations in 47 Tuc}
\shortauthors{Harbeck, Smith, \& Grebel}

\begin{document}


\title{CN  abundance variations on the main sequence of  47 Tuc\footnote{Based 
 on observations made with ESO Telescopes at the Paranal Observatory
 under programme ID 67.D-0153 and the MPG/ESO 2.2m telescope at La
 Silla Observatory.}}


\author{Daniel Harbeck}
\affil{Max-Planck-Institut f\"ur Astronomie}
\affil{ K\"onigstuhl 17, D-69117 Heidelberg, Germany}
\email{dharbeck@mpia.de}

\author{Graeme H. Smith}
\affil{Lick Observatory, University of California Observatories} 
\affil{University of California, Santa Cruz, CA 95064}
\email{graeme@ucolick.org}

\and
\author{Eva K. Grebel}
\affil{Max-Planck-Institut f\"ur Astronomie}
\affil{K\"onigstuhl 17, D-69117 Heidelberg, Germany}
\email{grebel@mpia.de}



\begin{abstract}
We report on a deep spectroscopic survey for star-to-star CN
variations along the main sequence (MS) of the globular cluster 47 Tuc
with ESO's VLT. We find a significant bimodal distribution in the
\scnb index for main-sequence stars in the mass range of $\sim
0.85$ to $0.65$~M\sunindex, or from the main-sequence turn-off down to
$\sim2.5$~mag below the main sequence turn-off.  An anti-correlation
of CN and CH is evident on the MS. The result is discussed in the
context of the ability of faint MS stars to alter their surface
composition through internal evolutionary effects. We argue against
internal stellar evolution as the only origin for the abundance spread
in 47 Tuc; an external origin such as pollution seems to be more
likely.
\end{abstract}


        

\section{Introduction}\label{Sect_Intro}

Globular clusters (GCs) are well suited to the study of ancient star
formation and the evolution of low-mass stars. In standard formation
scenarios all stars of a GC formed at the same time from a single
proto cloud; as a consequence all stars within a cluster should have
the same age and the same initial chemical composition. Indeed, within
many globular clusters the star-to-star iron abundance spread is
smaller than typical (spectroscopic or photometric) measurement errors
(e.g., as reviewed by \citealt{suntzeff93}, or in recent work by
\citealt{gratton01}).  Nevertheless, detailed studies have revealed
that there are significant star-to-star abundance variations of
certain light elements including sodium, aluminium, carbon, nitrogen,
and oxygen in globular clusters (e.g., the review by
\citealt{kraft94}).  One of the first discoveries of chemical
inhomogeneities in GCs was made by \citet{osborn71}, who found two red
giant stars in M10 and M5 to have enhanced CN molecular absorption
strengths compared to other stars in these clusters.

The first detailed studies of the abundance spread phenomenon
concentrated on the brightest stars at the tip of the red giant branch
(RGB); increasing telescope size and improved instruments have allowed
investigation of the chemical compositions of stars along the RGB down
to the subgiant branch (SGB) (e.g., \citealt{bell83,bell84};
\citealt{hesser84}; \citealt{smith89}; \citealt{suntzeff91}; 
\citealt{briley92}; \citealt{cohen02}). Even the main sequence
turn-off (MSTO) region of a few nearby GCs has been reached
\citep{hesser78,briley91,briley94,suntzeff91,cannon98,cohen99a,
cohen99b,briley01,gratton01}. The main result of all these efforts in
the last thirty years is the discovery of abundance variations in some
GCs along the whole stellar evolutionary sequence from the RGB to the
SGB and even to below the MSTO.  Some of the better studied examples
of globular clusters with abundance variations are 47 Tuc, M13, and
M3.

The ultimate tool for studying the chemical compositions of GC stars
is high-resolution spectroscopy of element absorption
lines. Unfortunately, for stars near the MSTO of globular clusters
this approach requires extremely long exposure times even with the
present day 10-meter-class telescopes.  However, the study of
absorption bands produced by molecules can serve as a useful tool to
investigate abundance variations even with moderate resolution
spectra.  For example, the broad CN and CH absorption features at
$3883$~\AA\ and $4300$~\AA\ respectively are measurable for faint
stars at the limits of large telescopes by either medium-resolution
spectroscopy or narrow-band photometry.

To investigate the origin of abundance inhomogeneities in globular
clusters, we have performed the deepest survey to date of CN band
absorption in main sequence stars of the globular cluster 47 Tuc. In
this paper we report on our observations with the Very Large Telescope
(VLT), which has resulted in the detection of significant star-to-star
differences in the CN abundance $2.5$~mag below the MSTO (i.e.
V=$21.5$~mag) of 47 Tuc. In Sections \ref{sect_variations}
\& \ref{sect_47tuc} we briefly review current scenarios that
could explain abundance spreads in GCs and describe earlier work on 47
Tuc. Sections \ref{sect_observation} \& \ref{sect_analysis} describe
our observations and the data analysis.  Our conclusions are presented
in Section~\ref{sect_conclusion}.

\section{Scenarios for Abundance Inhomogeneities in Globular Clusters}
\label{sect_variations}

Although the problem of abundance inhomogeneities, especially in the
CN molecule, within globular clusters was recognized over thirty years
ago, their origin is still unresolved. Over these thirty years three
scenarios have received considerable discussion in the literature: (i)
abundance changes produced internally by the evolution of GC stars,
(ii) GC self-pollution, and (iii) primordial inhomogeneities or
cluster self-enrichment. An excellent review of these scenarios can be
found in \citet{cannon98}, whose arguments we summarize here.

\subsubsubsection{Stellar Evolution}

If all stars in a globular cluster started with the same initial
chemical composition, the surface abundances of many of them evidently
have since changed by varying amounts during their subsequent
evolution.  In the stellar evolution scenario CNO-cycle processed
material from the interior of hydrogen-burning stars is dredged up
(e.g, by rotational mixing) to the stellar surface. The efficiency of
this dredge up is reflected by changes in the surface chemical
composition. This scenario can explain why chemical inhomogeneities
are generally found in the CNO elements, but not in the iron-peak
elements. Although according to conventional stellar evolution models
the dredge up of processed material happens only on the RGB, CN
variations are evident along the RGB, SGB, and at the MSTO of a number
of globular clusters such as 47 Tuc, M71, and M5, e.g.,
\citet{hesser78,hesser80,cannon98,cohen99a,cohen02}. As such, it is
not currently possible to account for the luminosity behaviour of CN
variations in GCs by a stellar evolution scenario. In many clusters
the CN band strengths at a given stellar luminosity show a bimodal
distribution (see e.g., \citealt{norris87,smith87}); the reason why
stellar evolution would routinely produce a bimodality rather than a
more uniform distribution is not yet understood.

\subsubsubsection{Self Pollution}

Asymptotic giant branch (AGB) stars with masses of $\sim 5$ M\sunindex
eject stellar winds that contain considerable CNO processed material
(e.g., \citet{ventura01}). Other types of stars such as planetary
nebulae progenitors or novae could also eject CNO-enriched
material. These stellar winds may be accreted by other stars,
including the present day MSTO, SGB, and RGB stars.  Stars that
accreted such material would show different surface compositions than
stars that accreted no material. If only the stellar surfaces were
polluted, deep convective mixing during the evolution along the RGB
should dilute the chemical inhomogeneities. This effect is not seen at
least in the case of the nitrogen abundance, although the [C/Fe]
abundance is found to decrease with luminosity on the RGB in some
clusters such as M92 and M13 \citep{carbon82, suntzeff81}.  Also, a
bimodal abundance distribution might not be expected from such a
phenomenon, but rather a range in the surface contamination.

\subsubsubsection{Primordial variations}

The assumption that all stars in a globular cluster had the same
chemical composition from the very beginning may not necessarily be
true. For instance, two molecular clouds of different chemical
composition could have merged and formed a globular cluster. If the
gas was not mixed, stars would reflect the chemical composition of one
or the other cloud. Here, a bimodal distribution could be
explained. An imperfect merging of the gas clouds would lead to a
range of abundances. Also, the first supernovae during the star
forming epoch of a GC may have polluted the parent molecular
cloud. Stars that formed after these events would be enriched in a
variety of elements. These two primordial scenarios may explain
abundance variations in the CNO elements, but they lack an easy
explanation of the chemical homogeneity of the iron-peak elements. In
a third ``primordial'' scenario two globular clusters experienced a
collision and merged after they had completed their star
formation. The result would be a bimodal abundance distribution, but
again the homogeneity of the iron-peak and other heavy elements is a
problem.

\section{The Globular Cluster 47 Tuc}
\label{sect_47tuc}

The globular cluster 47 Tuc is a very attractive object for a detailed
chemical investigation. It has a distance of only $4.5$~kpc and it is
after $\omega$ Cen the second-brightest cluster ($M_V=-9.4$~mag) in
the Milky Way. Due to a tidal radius of $47.25'$ it is quite an
extended object on the sky. The intermediate metallicity of the
cluster RGB stars ([Fe/H]=$-0.7$~dex) makes the formation of CN
molecules in their atmospheres more efficient than in metal-poorer GC
giants. Not surprisingly, a number of studies of CN abundance
variations in 47 Tuc have been carried out. In one of the earliest
studies \citet{mcclure74} found star-to-star differences in the DDO
C(41-42) CN index.  The distribution of CN band absorption strengths
at a given point on the RGB was shown to have a bimodal signature by
\citet{norris79}, with there being indications of an anticorrelation
between CN and CH band strengths \citep{norris84}. The main results
from studies in the last 30 years show CN variations along the RGB of
47 Tuc down to the MSTO. In the latest large photometric narrowband
filter survey of 47 Tuc, \citet{briley97} verified the existence of a
bimodal distribution of CN band strengths, with a radial gradient in
the relative fraction of stars with strong and weak CN bands; in the
inner part of the cluster there is a higher fraction of stars with
strong CN absorption than in the outer parts. The existence of this
gradient had been first noted by \citet{norris79}, and further
documented by \citet{paltoglou90}. References to additional studies
can be found in \citet{briley97}.

The deepest CN survey done in 47 Tuc so far (and in GCs in general)
extends to one magnitude below the MSTO \citep{cannon98}.  These
authors pursued deep spectroscopic work at the limits of a 4m-class
telescope. They showed that the bimodal distribution of CN band
strengths still exists on the upper main sequence. With the recent
availability of 8 to 10-meter-class telescopes and with the capability
of multi-object spectroscopy, large samples of stars on the fainter
main sequence now become accessible for chemical abundance
studies. The investigation of chemical inhomogeneities on the fainter
main sequence can contribute new evidence for or against the various
scenarios described above.

\subsection{New information on the main sequence}

In the evolutionary scenario of GC abundance variations all stars in a
cluster started with the same initial chemical composition; internal
evolutionary effects like deep mixing changed the star's surface
composition at a later date. To change the surface element abundances,
two mechanisms are required: (i) a nuclear process must change the
abundance of a certain element (in the case of the CNO process the
ratio of the elements C, N, and O would be altered), and (ii) some
interior mass transport mechanism(s) must dredge up the processed
material to the stellar envelope. In this study we concentrate on
abundance variations in the C and N elements since they can easily be
detected spectroscopically by variations in the CN bands.

In the low-mass stars on the main sequence of globular clusters the
pp-chain is the dominant process to produce nuclear energy in the
stellar core. In addition the CNO process is contributing to the total
energy output. Although the CNO cycle is a catalytic process, it can
alter the abundance ratios of the elements involved due to the
different time-scales of the various reactions in this
cycle. Especially the nitrogen content increases at the cost of carbon
and oxygen; the sum of C+N+O remains constant. The efficiency of the
CNO process depends strongly on the temperature; therefore this cycle
is the dominant nuclear process in young massive main sequence stars,
but its contribution to the energy output is marginal for stars of a
solar mass and below. According to stellar models globular cluster
stars at the upper main sequence with masses of $\sim 0.8$~M\sunindex
should not support deep mixing processes (the convective envelope
being relatively shallow at least in terms of the contained mass) so
that the stellar evolution scenario seems to fail to account for
observed CN variations at the MSTO. Additionally, mixing processes
would add fresh fuel to the hydrogen burning cores of the MS stars
that would affect the lifetime of their core-burning phase. As pointed
out in, e.g., \citet{cannon98}, no such effect can be seen in 47
Tuc. Nevertheless, the understanding of deep mixing or other material
transportation processes is still incomplete and other mechanisms like
diffusion should still be taken into account.

\begin{figure}
\plotone{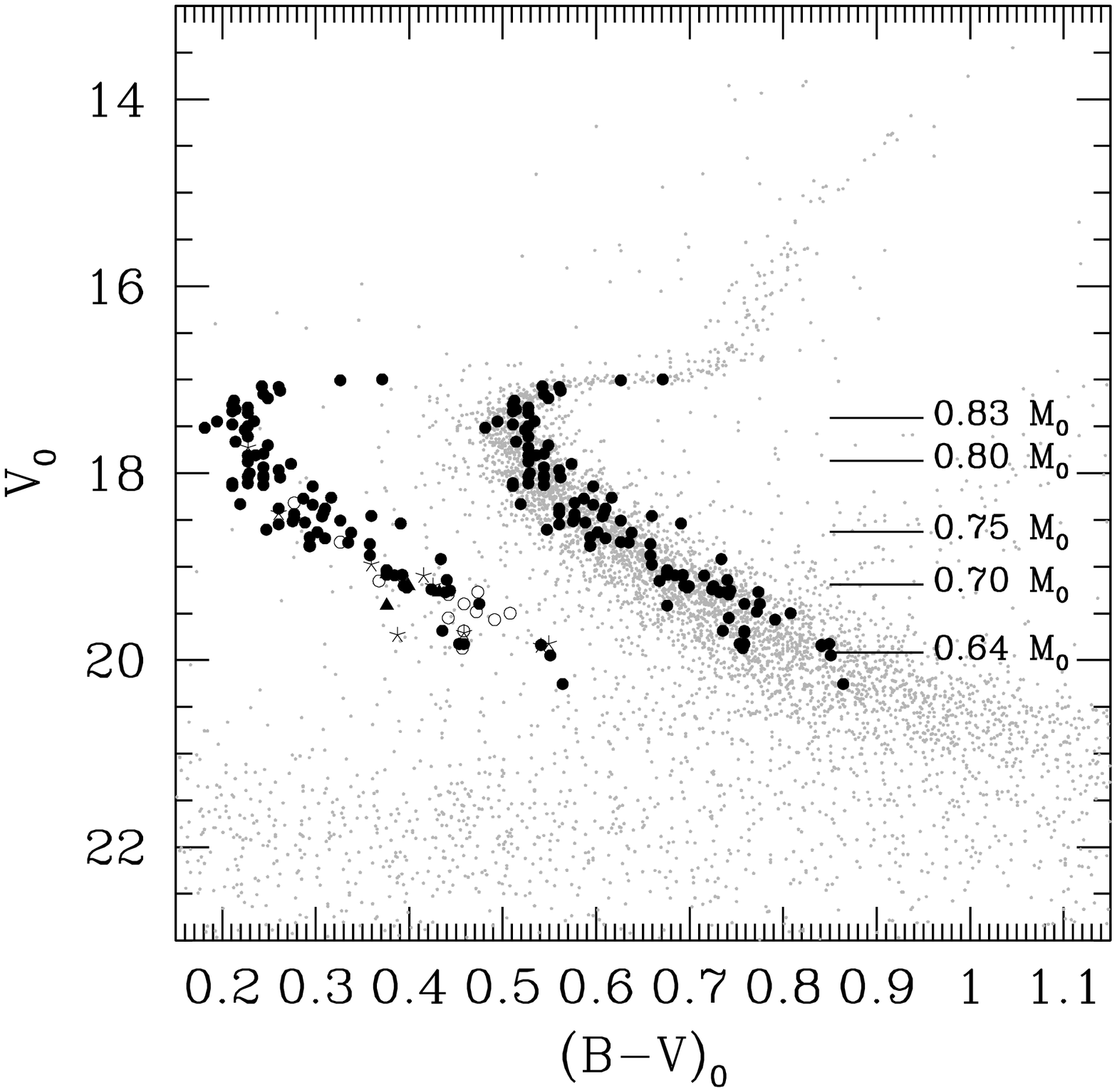}
\caption{Color
  magnitude diagram of 47 Tuc based on WFI observations. All program
  stars are plotted on the main sequence with filled dots. Based on
  isochrones for $Z=0.004$ and an age of 14~Gyr \citep{girardi00} we
  show the stellar masses at various points along the main
  sequence. For better visibility, all program stars are plotted again
  with an offset of $0.3$~mag in (B-V). Stars with clean CN
  measurements are plotted with filled circles (quality=''0''), while
  ambiguous measurements (quality= ``1'') are indicated by open
  circles. Four cluster non-members, that are possibly SMC giants, are
  marked with triangles. Program stars with unreliable spectroscopy
  are marked by asterisks (quality=``2''). \label{fig_cmd}}
\end{figure}

Turn-off stars in 47 Tuc have masses of $0.85$ M\sunindex (see
Fig~\ref{fig_cmd}).  \citet{cannon98} showed that stars at the
turn-off and below down to a magnitude of $V=18.6$~mag or a mass of
$\sim 0.75$~M\sunindex exhibit CN abundance differences.  In our
study, we have measured stars with magnitudes down to $V=20$, or 2.5
mag below the MSTO, which have masses around $0.65$~M\sunindex
(Fig.~\ref{fig_cmd}). The (model-dependent) core temperature of these
stars is $\log(T_c)\sim 7.10$ at the MSTO and $\log(T_c)\sim 7.01$ for
the fainter stars. These stellar parameters were taken from isochrones
for a metallicity of $Z=0.004$ and an age of $14$~Gyr
\citep{girardi00}.  Due to the strong dependence of the CNO cycle
energy production rate $\epsilon_{CNO}$ on the temperature
($\epsilon_{CNO} \propto T^{17}$), the contribution of the CNO cycle
relative to the energy production by the pp-chain ($\epsilon_{pp}
\propto T^{6}$) will be reduced by a factor of approximately ten in
the fainter stars. This estimate is based on the decreased core
temperature only and does not account for a decreasing core
density. Therefore the CNO cycle efficiency in stars $2.5$~mag below
the MSTO is suppressed by {\it at least} this factor of ten. If
hydrogen-burning products could be dredged up to the surface, the
amount of CNO processed material that could lead to CN variations
would also be reduced by a factor of at least ten for the fainter
stars.

This simple calculation can be summarized in the following statement:
along the main sequence the production rate of N due to the CNO cycle
becomes completely suppressed for the lower mass stars with masses of
$\sim 0.65$~M\sunindex. If there is still some scatter in the CN
content between faint stars along the MS, internal stellar evolution
cannot explain this scatter because there is no internal CNO-process
working to change the ratio of the CNO elements, and external
processes (outside a given star) must have played an important
role. If the abundance spread disappears among the fainter stars, this
would be clear evidence for internal mixing processes as the origin of
the CN abundance riddle. Measurements of the CN content of main
sequence stars can therefore distinguish between the internal
evolutionary scenario of CN variations and external
scenarios. However, CN measurements alone cannot distinguish between
different external scenarios. The main sequence study of
\citet{cannon98} extends to a mass of $\sim0.75$~M\sunindex.  At
this mass, the CNO process efficiency is reduced by only a factor of
$\sim3$.  Surveying CN bands at even lower masses can greatly
strengthen the constraints placed on the stellar evolution scenario by
way of the above CNO-cycle efficiency argument.

\section{Data and Reduction}
\label{sect_observation}

\subsection{Observations}

\subsubsubsection{Target selection from WFI imaging}

Candidate stars in 47 Tuc for spectroscopy with the VLT of the
European Southern Observatory (ESO) at Cerro Paranal, Chile, were
chosen from CCD images of the cluster.  In September 2000 we observed
47 Tuc with the MPG/ESO 2.2-m telescope at La Silla, Chile, using the
Wide Field Imager (WFI) \citep{baade98} with Johnson B and V
filters. The exposure times were $3 \times 300$~s in each
filter. Conditions were non-photometric with a seeing of $1.5''$. The
WFI offers a field of view (FoV) of $34\time33$~arcmin$^2$ and
consists of a mosaic of eight $4096\times2048$ CCDs. The WFI was
centered on the globular cluster. The radius of the FoV is
$\sim0.8^\circ$, which is approximately half the tidal radius of 47
Tuc. The basic processing of the raw CCD frames was done using the
tasks {\tt esowfi} and {\tt mscred} in IRAF\footnote{IRAF is
distributed by the National Optical Astronomy Observatories, which are
operated by the Association of Universities for Research in Astronomy,
Inc., under cooperative agreement with the National Science
Foundation.}. To facilitate proper sky subtraction and to reduce the
incidence of blended images we decided to only select stars in the
outer, less-crowded regions of 47 Tuc for the follow-up
spectroscopy. We therefore concentrated on chip \#1 of the CCD mosaic
(at the north-east corner of the FoV) to define the sample of
spectroscopy candidates. With chip \#1 we sample a radial coverage
from $8'$ to $25'$, or $0.3$ to $0.5$ tidal radii. For the selected
chip we performed PSF photometry using the DAOPHOT \citep{stetson92}
implementation in IRAF.  Since the observations were obtained under
non-photometric conditions and no local standard stars are available
in the area covered by chip \#1, we matched the observed MS and the
SGB to fiducial lines published by \citet{hesser87}.  The calibrated
photometry was dereddened assuming $E(B-V)=0.04$ \citep{harris96}. The
resulting, dereddened color-magnitude diagram is shown in
Fig.~\ref{fig_cmd}.

\subsubsubsection{Spectroscopy with the VLT/FORS2/MXU}

The goal of our spectroscopic observations was to measure the
strengths of the $3883$~\AA\ and $4215$\AA\ CN absorption bands. We
used the Focal Reducer and Low Dispersion Spectrograph 2 (FORS2) at
the ESO VLT.  FORS2 is equipped with a mask exchange unit (MXU) which
allows multi-slit spectroscopy with freely defined metal slit masks
and a FoV of $6.8\times6.8$~arcmin$^2$. We defined three slit masks
with the FORS instrument mask simulator (FIMS) software package
provided by ESO. The positions of the program stars were determined on
CCD images that were obtained at the FORS/MXU. The exposure time for
the preimages was $45$ seconds with a V filter. The slit width on the
masks was fixed to $1\,''$ and the slit length was chosen to be at
least $8\,''$ to allow for local sky subtraction. Typically 35 slits
were fitted onto one mask.  In some cases two stars could be observed
with one slit. We selected the grating B600 with a dispersion of
$50$~\AA\,mm$^{-1}$ or $1.2$~\AA\,pixel$^{-1}$. In combination with
the $1\,''$ slit width this results in a nominal spectral resolution
of 815, or a line width of $\le5$~\AA. The actual spectral coverage
depends on the location of a star on the mask, but we took care to
cover at least the range $3800$~\AA\ to $5500$~\AA. All program stars
are listed in Table~1. The observations were carried out in service
mode between 2001 July 18 and 2001 July 26. Each mask configuration
was observed four times with exposure durations of $2925$~s each,
leading to a total exposure time of 3.25 hours per mask. The seeing
conditions were almost always better than $1\,''$ according to our
specification for the service observations. Additionally, bias, flat
field, and wavelength calibration observations were obtained.

\subsection{Data reduction}

All data were reduced using the {\tt ccdred} and {\tt specred}
packages under IRAF. First all frames were overscan and bias
corrected. The flat fields for each night and mask were stacked.
Since each slit mask was observed four times and the alignment of the
frames was excellent, we coadded the four frames for a given slit mask
with cosmic ray rejection enabled. The resulting frames were almost
free of cosmic rays. For further analysis we extracted the area around
each slit and treated the resulting spectra as single-slit
observations. The wavelength-calibration images and flat fields were
treated in the same manner. We fitted a tenth-order polynomial to the
flat field images to correct for the continuum. The (separated) object
spectra and arc images were flat-field calibrated with these
continuum-corrected flat fields.  The {\tt doslit} task was used to
wavelength calibrate, sky-subtract, and extract the stellar
spectra. The wavelength solution from the HgCd arcs was fitted by a
fourth-order spline. The typical rms of the wavelength calibration is
on the order of $0.2$\AA, which is expected at the given spectral
resolution.

\subsubsubsection{Spectral indices}

We measure the strength of the CN absorption bands at $3883$~\AA\ and
$4215$~\AA\ using spectral indices \scnb and \scnr similar to those of
\citet{norris79,norris81}.  These indices basically quantify the CN
content of the atmospheres of red giant stars. Unfortunately, the main
sequence stars we are considering here have strong hydrogen lines in
the region of the \scnb spectral index. We therefore modify this index
according to \citet{cohen99a} to exclude these hydrogen
lines. Additionally we quantify the CH molecular absorption at
$4300$~\AA\ using a CH(4300) index defined according to Cohen
(1999a,b). As a check we additionally extract an HK index that
measures the absorption strength of the Ca\,{\sc ii} H+K lines. The
index definitions are:

\begin{eqnarray}
\rm{S}(3839) & = & -2.5 \log{\frac{F_{3861-3884}}{F_{3894-3910}}} \\
\rm{S}(4142) & = & -2.5 \log{\frac{F_{4120-4216}}{0.4\,F_{4055-4080} 
                                  +0.6\,F_{4240-4280}}} \\
\rm{CH}(4300)& = & -2.5 \log{\frac{F_{4285-4315}}{0.5\,F_{4240-4280}
                                  +0.5\,F_{4390-4460}}} \\
\rm{HK}      & = & -2.5 \log{\frac{F_{3910-4020}}{F_{4020-4130}}} 
\end{eqnarray}

where $F_{3861-3884}$ for example is the summed spectral flux (in our
case ADU counts) from $3861$\AA\ to $3884$\AA. Each index quantifies
the absorption strength of a molecular band or atomic lines relative
to a nearby pseudo-continuum. For stronger band absorption the index
values increase.

\begin{figure}
\plotone{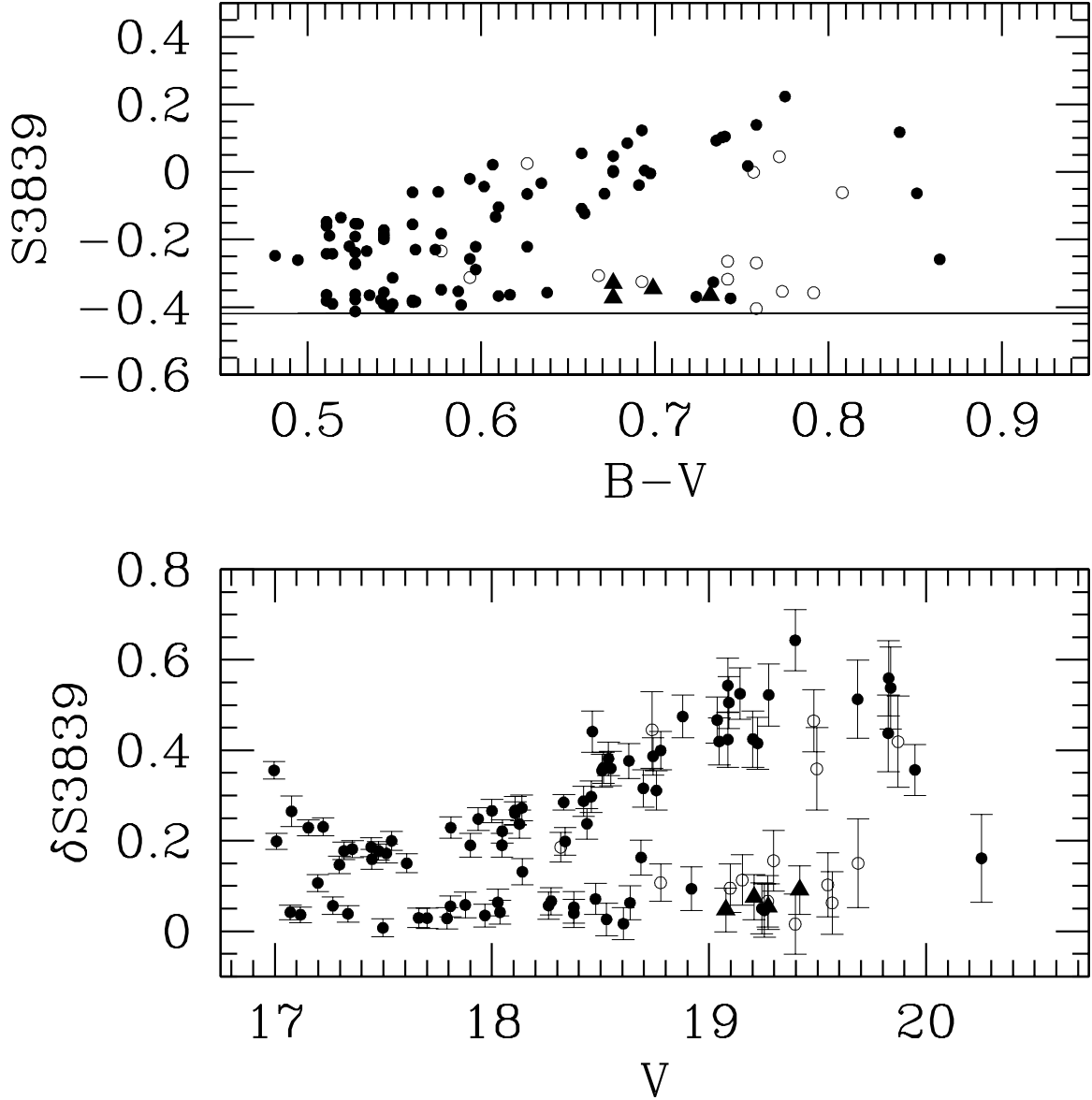}
\caption{The CN index \scnb
  plotted against the B-V color of 47 Tuc main sequence stars (upper
  panel). We correct for a zeropoint such that CN-weak stars have a
  corrected index $\delta \rm{S}(3839)=0$. This corrected CN index is
  plotted in the lower panel versus the stellar V magnitude. Note the
  significant scatter in the $\delta$\scnb index even for the faintest
  stars. For stars fainter than $V=17.5$~mag the separation between
  CN-strong and CN-weak stars is more than 0.2 in \scnb and increases
  with inreasing magnitude. The distribution of CN-poor and
  CN-enhanced stars does not seem to be random but is bimodal. Symbols
  are the same as in Fig.~\ref{fig_cmd}.  \label{fig_S3839}}
\end{figure}

\begin{figure}
\plotone{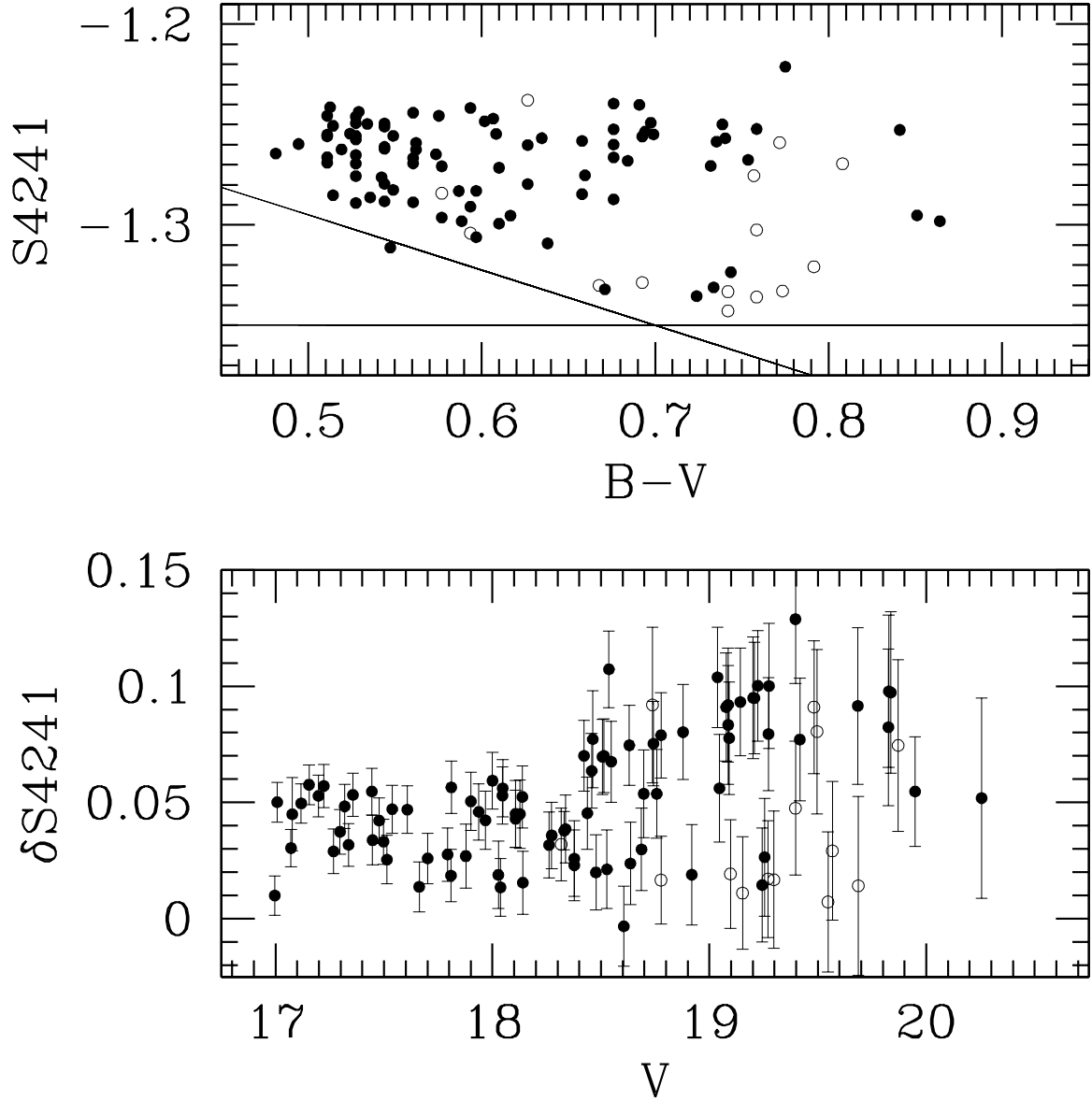}
\caption{Same as Fig.~\ref{fig_S3839}, 
        but for the weaker CN band index \scnr. Since a baseline to
        the \scnr index exhibits a stronger dependency on the
        temperature we correct with a linear fit (the pair of solid
        lines).  Note the increased bimodality for the fainter
        stars. Symbols are the same as in
        Fig.~\ref{fig_cmd}. \label{fig_S4241}}
\end{figure}

The derived indices \scnb and \scnr are plotted versus the $(B-V)$
color of each star in the upper panels of Fig.~\ref{fig_S3839} and
Fig.~\ref{fig_S4241}. The formation efficiency of the CN molecule
depends on the temperature, and therefore the indices can have a
dependence on the color of the stars. For \scnb the effect here is
negligible. To correct for this effect in the \scnr index, we assume a
combination of two straight lines (see Fig. \ref{fig_S4241}) as a
lower envelope and subtract this fit from the data. A constant value
(see Fig.~\ref{fig_S3839}) is subtracted from \scnb such that the
corrected values approximately equal zero for the CN-poor stars. The
corrected CN indices will be referred to as $\delta$\scnb and
$\delta$\scnr; they are plotted versus the stellar V magnitudes in the
lower panels of Fig.~\ref{fig_S3839} and Fig.~\ref{fig_S4241}. The
derived index values can be found in Table~\ref{tab_stars}. The
plotted error bars are calculated assuming Poisson statistics in the
flux summations.

To verify the reliability of the \scnb index we visually classified
the CN absorption strength for all stars into five classes.  We found
an excellent correlation between the \scnb index value and our visual
classification. We have assigned a "quality" flag to the data for each
star. Stars for which spectra show clean CN absorption bands with few
noise imperfections and for which a reasonably confident CN
classifications was felt to be obtainable have been assigned a data
flag of ``0''.  In some cases, however, we could not judge a unique CN
classification by eye, especially for a few faint stars near to the
center of 47 Tuc. Such instances tend to occur when spectra have a low
S/N or do not show consistent trends in the three main absorption
features at 3860\AA,3875\AA, or 3889\AA\ of the blue CN band. These
stars may be blended with others (as suspected by their proximity to
the cluster center) and are flagged with a ``1'' in
Table~\ref{tab_stars}. Nevertheless, the stars in this quality class
generally follow a relation between the visual classification and the
measured \scnb absorption strength.  For a few stars the data
reduction produced poor S/N spectra. This may probably be due to a
misallignment of the slits. These stars are flagged by a ``2'' in
Table~\ref{tab_stars} as poor quality spectra. In
Figs.~\ref{fig_S3839} and \ref{fig_S4241} stars for which the spectrum
obtained is assigned to quality class ``1'' are plotted with open
circles. Stars that did not permit visual classifications are not
plotted.

\subsection {Membership of program stars}

\subsubsubsection{Field star contamination}

We estimate the expected number of galactic foreground stars in our 47
Tuc field. According to \citet{ratnatunga85} the expected
contamination by the Milky Way is $\sim 3.5$ field main sequence stars
per arcmin$^2$ in a $2$~mag interval from $V=18$ to $V=20$ and a
$0.5$~mag interval in $(B-V)$ centered on the main sequence
fiducial. We selected our spectroscopy candidate stars from a
$0.05$~mag wide region in (B-V) about the 47 Tuc main sequence over a
luminosity range from $V \sim17.5$~mag to $\sim 20$~mag. We used only
a small fraction of the FoV of FORS for candidate star selection due
to our contraints on the wavelength coverage. Per mask configuration
an effective FoV is $\sim6'\times 2'=12$~arcmin$^2$. The expected
number of field stars in the field of view of our three slit masks is
therefore $\case{3.5\,\rm{stars/arc min}^2}{\cdot
2\,\rm{mag}\cdot0.5\,\rm{mag}} \cdot 2.5\,\rm{mag}
\cdot 0.05\,\rm{mag}
\cdot 3\,\rm{masks} \cdot 
12 \, \rm{arc min}^2 \sim 16\,\rm{stars}$. The number of stars
observed in the FoV of three slitmasks and the selected part of the 47
Tuc main sequence is $\sim 2500$. In our sample of $115$ spectroscopic
survey stars we therefore expect a total number of
$\case{115}{2500}\cdot 16 \sim 1$\ field star. We conclude that the
contamination by galactic field stars that cannot be identified by
their radial velocity has a negligible effect on our results and does
not account for the CN bimodality seen within our sample of stars.

\subsubsubsection{Contamination by the SMC}
\label{sect_smccont}

In the color magnitude diagram of 47 Tuc (Fig.~\ref{fig_cmd}) the RGB
of the Small Magellanic Cloud (SMC) crosses the main sequence of 47
Tuc at a magnitude of $V \sim 20$~mag. The SMC has a radial velocity
of $+163$~km\,s$^{-1}$, while the radial velocity of 47 Tuc is $-18$
km\,s$^{-1}$.  We determined the radial velocities (RVs) of the
observed stars by measurement of the center of the calcium H+K lines
as well as the H$\delta$ absorption line. The typical rms of the RV
measurement for a single star is of order $30$~km\,s$^{-1}$ to
$40$~km\,s$^{-1}$. For the well-sampled spectra we expect an accuracy
of $\sim \case{c}{850}\times \case{1}{10}=35$~km\,s$^{-1}$, which is
in good agreement with the actual measured scatter.  The {\tt
rvidlines} task in the {\tt rv} package in IRAF was used to determine
the centers of the absorption lines.  We found four stars with notably
different radial velocities from the rest of our sample; they have
magnitudes from $V=19$~mag to $V=20.25$~mag. These stars (number 3206,
4592, 5296, and 6335 in Table 1) could be SMC members from their
location in the CMD, but since they are among the fainter stars in our
sample the measured RVs could be affected by errors.

\subsection {A correlaton between CN absorption bands}

\begin{figure}
\plotone{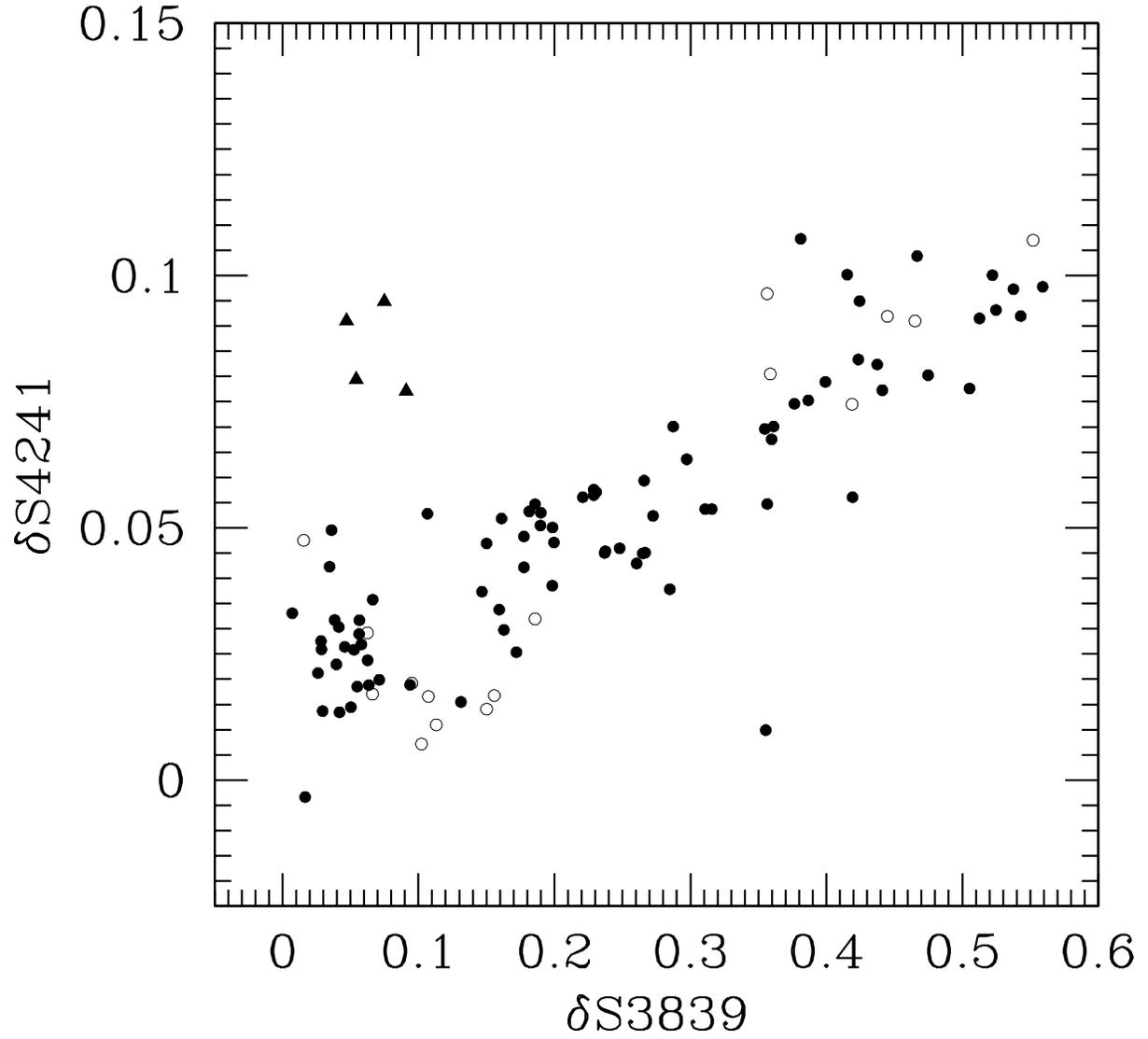}
\caption{Comparison of
  the CN indices $\delta$\scnb and $\delta$\scnr. Both indices, which
  measure absorption by the same molecule, are
  well-correlated. Symbols are the same as in Fig.~\ref{fig_cmd}.
  \label{fig_compcn}}
\end{figure}

As a diagnostic we plot $\delta$\scnb vs. $\delta$\scnr in
Fig.~\ref{fig_compcn}. Again, stars having spectra with some
ambiguities (quality class ``1'') are plotted with open circles. The
outstanding group of four stars at $\delta$\scnb$ \sim 0.08$ and
$\delta$\scnr$ \sim 0.08$ (plotted with filled triangles) are the
stars noted in Section~\ref{sect_smccont} as having radial velocities
well off the measured RV distribution of 47 Tuc and are probably not
cluster members. Although there is scatter present, the correlation
between the two CN indices stands out. The star at
$\delta$\scnb$=0.35$ and $\delta$\scnr$=0.01$ that does not follow the
correlation is the reddest SGB star (compare to
Fig.~\ref{fig_cmd}). Since the $\delta$\scnr index is less sensitive
to CN abundance, and in any event correlates with $\delta$S(3839), we
will concentrate on the $\delta$\scnb index in the remainder of this
study.

\section{CN variations on the main sequence}
\label{sect_analysis}

\begin{figure}
\plotone{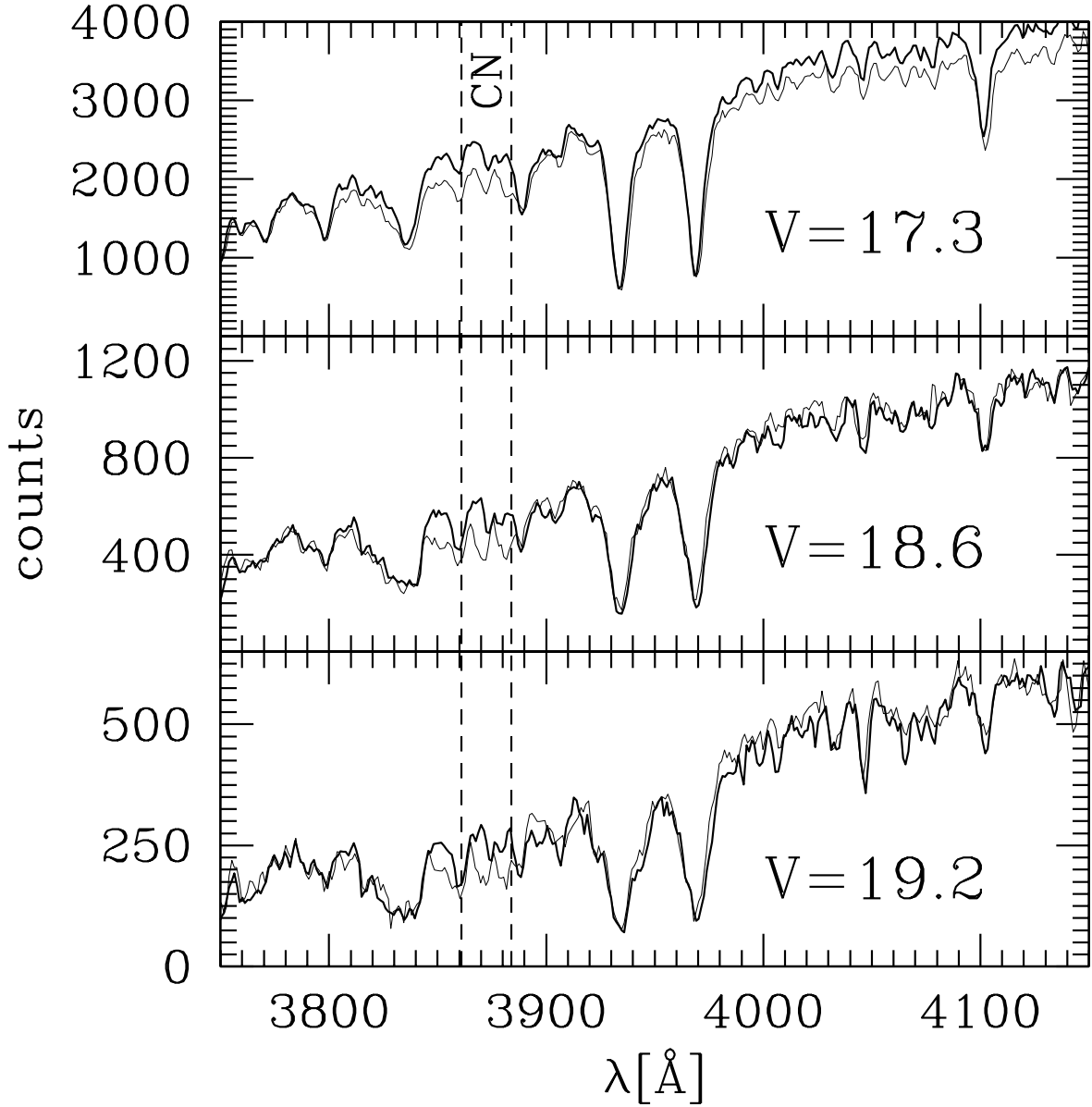}
\caption{Demonstration 
 of the different CN absorption band strengths at 3883\AA\ (between
 the dashed lines). We show example spectra of CN-strong (thin lines)
 and CN-weak (thick lines) stars at three different V magnitudes.  The
 stars plotted are \#806, \#3060, \#4344, \#5545, \#5759, and \#5809.
 \label{fig_real}}
\end{figure}

In the lower panel of Fig.~\ref{fig_S3839} the $\delta$\scnb index is
plotted versus the stellar V band magnitudes. A pronounced variation
in the measured CN index across the whole luminosity range covered is
obvious. Even more, the distribution of CN absorption strength is
bimodal: stars seem to be either CN-strong or CN-weak. There might be
a possible turn-over at the faint end of our sample, although this
could be an effect of increasing measurement errors. Our data
demonstrate that CN variations exist in 47 Tuc even 2.5~mag below the
MSTO. The range of the CN variations becomes more evident for lower
luminosities ($V\ge18.5$), which is probably an effect of the
decreasing surface temperature and therefore increased efficiency of
molecule formation. \citet{rose84} showed for a sample of nearby dwarf
stars that from spectral type G0 to K the CN 3883 band strength
steadily increases, with the metal-richer stars showing a steeper
increase as a function of spectral type. The increasing amplitude of
the bimodal CN distribution is also evident in the less sensitive
$\delta$\scnr index in Fig.~\ref{fig_S4241} (lower panel). Note that
this effect was not seen in the study of \citet{cannon98}, which only
extends to a magnitude of $V=18.6$~mag. At this luminosity the
bifurcation of the CN band strength starts to increase. In
Fig.~\ref{fig_real} we show spectra of CN-strong and CN-weak stars in
three luminosity bins. Differences in CN band absorption are clearly
visible, which was also seen in Fig.~\ref{fig_S3839}. We wish to point
out here that at the luminosity of V$\sim 19.25$ the CN-weak sample
suffers from SMC field star contamination. The two faintest stars with
good quality spectra have luminosities of $V=19.25$~mag. CN-weak stars
fainter than this luminosity still have reliable spectra, albeit of
more limited S/N. Nevertheless, a verification by additional
observations appears appropriate.

\subsection{A Test of the CN Bimodality}

\begin{figure}
\plotone{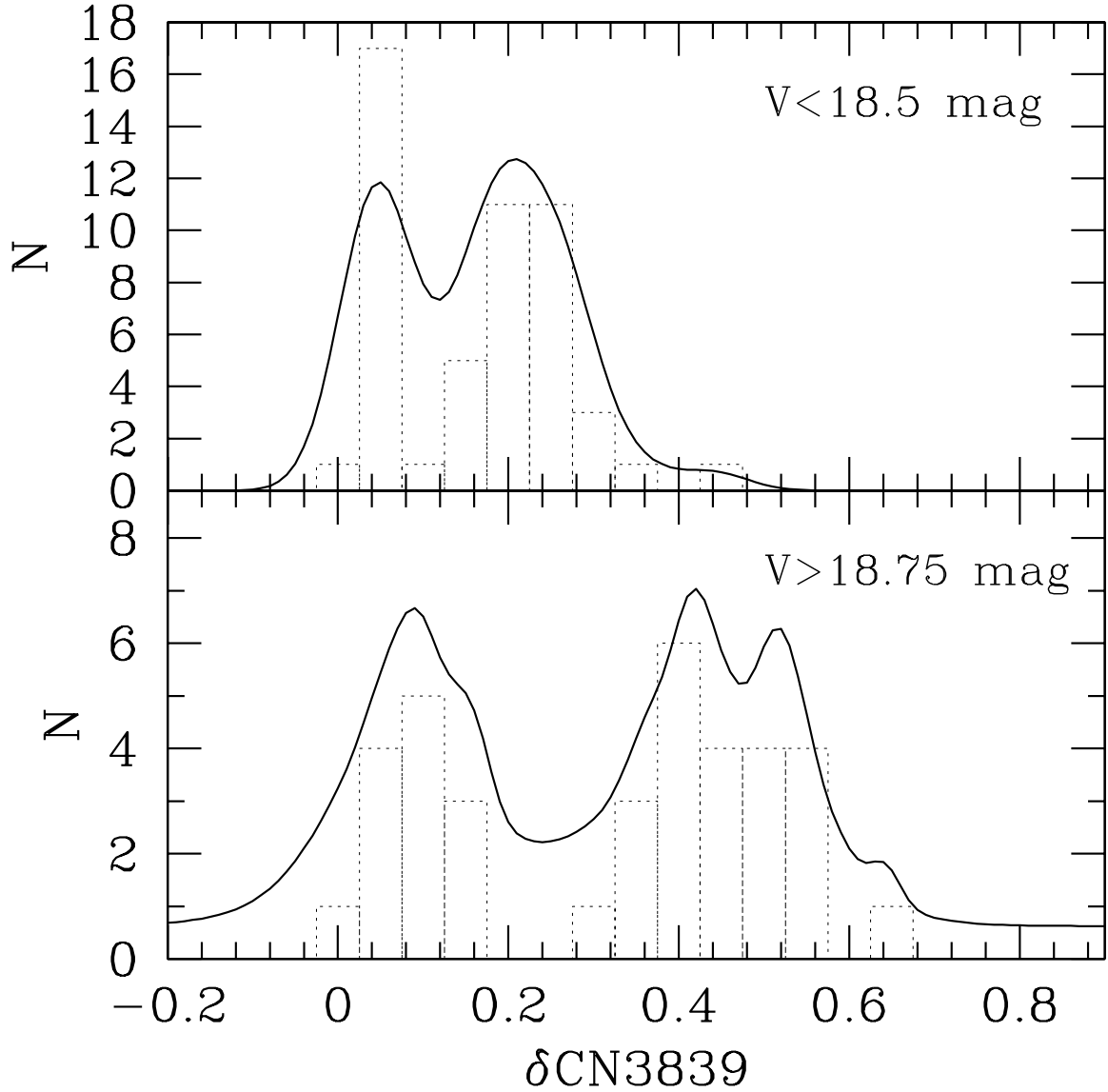}
\caption{Distribution of 
 the CN absorption strengths $\delta$\scnb for stars brighter than
 $V=18.5$~mag (upper panel) and fainter than $V=18.75$~mag (lower
 panel). The histogram bars show absolute number counts. The thick
 line is an error-folded density plot of the $\delta$\scnb
 distribution. The bimodality of the distributions stands out in both
 sub-samples. \label{fig_cndist}}
\end{figure}

The bimodality of the CN absorption strength is an outstanding feature
of the main sequence of 47 Tuc. A bimodal CN distribution has been
reported previously for the RGB, the SGB and the turn-off region
(e.g., \citealt{norris79}, \citealt{cannon98}, and references
therein). In Fig.~\ref{fig_cndist} we plot histograms of the CN band
strength $\delta$\scnb for magnitude ranges $V\le 18.5$~mag and
$V\ge18.75$~mag. The thick line is a smoothed plot of the
$\delta$\scnb values, where each measurement was folded by a Gaussian
having a width equal to the error bar.  The division into two
luminosity bins is motivated by the increased absorption strengths of
CN-enriched stars fainter than $V=18.5$~mag. A histogram of the CN
absorption strength of all stars would smear out the bimodality due to
the different range in index values at different magnitudes.

We quantify the significance of this bimodality with a {\tt kmm}-test
\citep{ashman94}. For the bright and faint
samples we obtain confidence levels of $99.996$~\% and $99.999$~\%
respectively for the distributions to be bimodal. This is very
significant (indeed, the significance for the faint sample reached the
numerical accuracy of the program used to calculate it).

\begin{figure}
\plotone{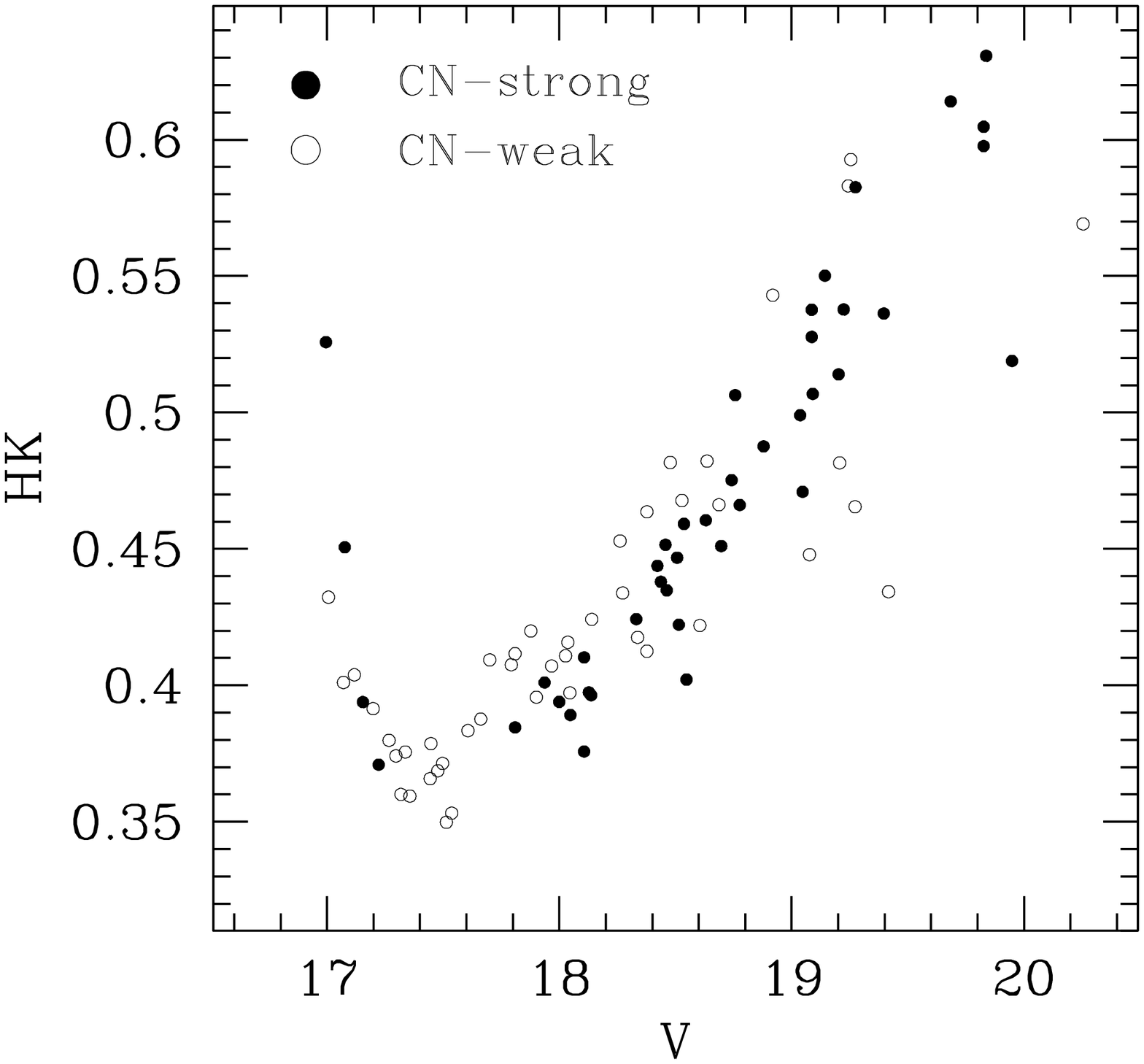}
\caption{ Distribution 
   of Ca\,{\sc ii} H+K absorption strength as a check of the data
   reduction. No bimodality such as for the CN distribution is seen,
   but a very weak anti-correlation between the CN and HK
   absorption. \label{fig_hkplot}}
\end{figure}

In order to double-check if the bimodal distribution of CN strength
could be caused by observational effects or errors in the data
reduction, we plot the Ca\,{\sc ii} H+K index defined in equation (4)
versus the stellar V magnitude in Fig.~\ref{fig_hkplot}. Higher values
of HK indicate stronger calcium absorption. No bifurcation as for the
CN indices is apparent. In this plot we mark CN-rich and CN-poor stars
with filled and open circles, respectively.  Although there is no
clear trend, CN-poor stars might tend to have higher HK absorption and
vice versa, but this effect is within the errors of the index
measurements. We conclude that there is no obvious effect in the data
that could mimic the bimodal distribution in the CN absorption
strengths.

\subsection{The CN/CH anticorrelation}

In 47 Tuc an anti-correlation between CN and CH absorption band
strength was reported in earlier studies of red giants
(\citealt{norris84}; \citealt{briley94}, and references therein). Such
an anti-correlation indicates that an enhancement of CN coincides with
a decrease in CH, or with a depletion of carbon at the expense of
nitrogen. Such an anti-correlation is expected if CN variations are
caused by variable amounts of CNO cycle processing. If the CNO process
only alters the ratio of the CNO elements, the sum C+N+O must stay
constant. An increase of nitrogen at the cost of carbon would (to a
certain limit) increase the CN abundance but decrease the absolute
content of carbon. As a result, the formation of the CH molecule would
be suppressed.

\begin{figure}
\plotone{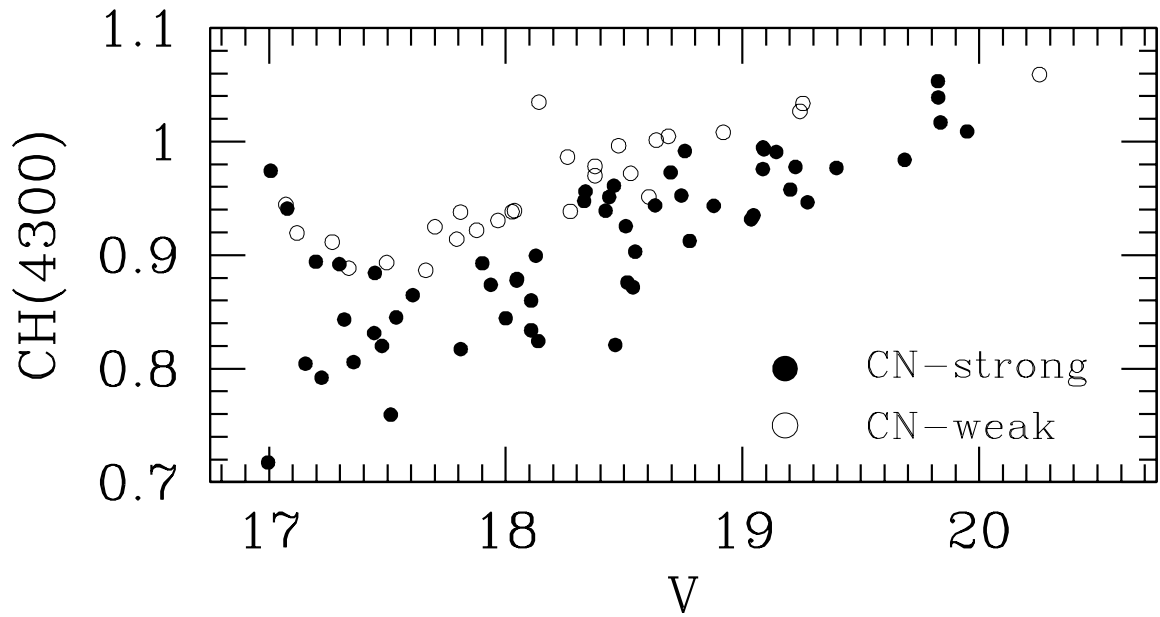}
 \caption{CH(4300) index plotted vs. the stellar magnitude for all
 stars with clean CN measurements (quality = ``0''). Stars classified
 as CN-strong are plotted with filled circles; CN-weak stars are
 plotted with open cirles. Note that all CN-weak stars have strong CH
 absorption. \label{fig_cnch}}
\end{figure}

In Fig.~\ref{fig_cnch} we plot the measured \sch\ index versus the V
band magnitude for all stars with good spectra (quality flag =
``0''). Stars classified as CN-strong are plotted with filled circles,
CN-weak stars are drawn with open circles. It is evident that stars
with the strongest CH bands are weak in CN absorption. For the
faintest stars in the sample ($V\ge19.5$) this anti-correlation does
not stand out clearly; we find that increasing absorption line
strengths for these cooler stars affects the measurement of the
comparison regions in the CH index.

\section{Summary and Conclusions}
\label{sect_conclusion}

Bimodal CN variations are detected in 47 Tuc among main sequence stars
with masses down to 0.65 M\sunindex or $2.5$~mag below the MSTO. This
extends earlier studies to stars that are nearly $1.5$ magnitudes
fainter.  There is an anticorrelation between the CN and CH absorption
strengths on the main sequence of 47 Tuc.

\subsection{External vs. evolutionary scenarios}

In Section 3.1 we argued that for stars $2.5$~mag below the MSTO the
CNO-cycle contribution is reduced by at least a factor of ten relative
to the MSTO. Despite the fact that the faint stars in our sample
should not be able to alter their chemical composition by themselves
(by CNO-cycling plus mixing processes), they still show clear and
significant star-to-star CN variations. Evidently the internal stellar
evolution scenario seems be ruled out, at least as the {\it only}
explanation for the CN-variation phenomenon in 47 Tuc. It is important
to mention that from the tip of the RGB to the faint main sequence
both (i) a bimodal CN distribution, and (ii) a CN/CH anti-correlation,
are found in 47 Tuc. This suggests that the same effect causes the
abundance spread from the tip of the RGB to the faint main sequence.

A caveat of our results is that an increased CN formation efficiency
in the cooler atmospheres of the faint MS stars may compensate for any
vanishing scatter in the CN content. The greater separation between
CN-strong and CN-weak stars at the faint end of our sample in
Fig.~\ref{fig_S3839} may indicate that this effect is important.  We
plan to investigate temperature effects on CN formation in our sample
by model atmosphere calculations in a subsequent paper. Additional
observations would help to make our result more secure: for stars just
one magnitude fainter than our sample the efficiency of the CNO-cycle
relative to the pp-chain drops down by another factor of ten. Any
diminution in the effects of internal evolution on the behaviour of CN
would be even more evident among such stars than in our current study.

If internal stellar evolution can be ruled out, external scenarios
remain to be discussed; either self-enrichment during the star forming
episode of 47 Tuc or later self-pollution. However, early
self-enrichment by high-mas stars is unlikely to produce only an
imprint in the lighter elements like CNO. Self-pollution by high-mass
AGB stars on the other hand could explain the lack of abundance
variations in the heavy elements (e.g., as discussed in
\citet{cannon98}). However, the bimodality of the CN band
strength still needs to be explained, and the AGB-star
enrichment/pollution scenario is not without other challenges
\citep{denissenkov97}.

\subsection{On the origin of the bimodality}

Current scenarios can hardly explain all aspects of the bimodal
distribution of CN absorption strengths in 47 Tuc or in other GCs.
For example, it is not clear how internal pollution by AGB winds would
lead to a bimodal abundance distribution. The measured bimodal CN
distribution does not necessarily reflect a bimodal abundance spread:
the distribution of CN absorption strengths reflects the true
distribution of atomic abundances folded with the curve of growth
(COG). It might be that all CN enhanced stars in 47 Tuc are polluted,
but they would fall on the flat part of the COG. A continuum of atomic
carbon and nitrogen abundances could therefore lead to more or less
the same CN absorption strength at a given effective temperature. Such
a ``saturation'' of the CN S(3839) band was proposed by
\citet{suntzeff81} and \citet{langer85} to explain the bimodality of
CN absorption in GC RGB stars. The similarity of CN absorption
strength for the CN-rich stars might be explained by this saturation
scenario, but the homogeneity of CN strength of the CN-poor stars
remains unsolved: These stars would fall on the raising part of the
COG; if pollution leads to a spread of abundances one would expect a
continuum of CN absorption strength until the flat part of the COG is
reached. This is not seen in the bimodal CN distribution of 47 Tuc. If
indeed saturation happens for the CN-strong stars, then the CN-weak
stars may represent the unpolluted, chemically homogeneous fraction of
the cluster whose CN abundance falls on the raising part of the
COG. Due to their chemical homogeneity they would have the same CN
absorption strengths at a given effective temperature. The bimodal
pollution mechanism still remains to be explained.

Binarity combined with AGB star pollution might be a solution: Only
stars in binary systems would have experienced pollution by accreting
the stellar winds of their evolving higher mass AGB companion. If
single stars accreted only a negligible fraction of intracluster gas
during the AGB contamination phase, the bimodality in the CN
distribution would simply reflect the binary fraction in the globular
cluster. If pollution in binary systems played an important role in
GCs, one might also expect to find a significant number of CH
stars. However, despite intense searches only a handful of CH stars
have been found so far (e.g., \citealt{cote97} and references
therein).

\subsection{Where is the polluting gas?}

If self-pollution happened in globular clusters, should the polluting
gas still be visible? The present-day amounts of gas in GCs appear to
be extremely small and below the detection limits of classical
approaches. Recently, ionized gas was detected in 47 Tuc from
the measured radio dispersion of millisecond pulsar signals
\citep{freire2001}.  The authors estimate the total gas mass in the
cluster's center to be roughly 0.1 M\sunindex. This amount is only a
thousandth of the expected gas loss from stellar winds that should be
ejected into the cluster between two passages through the Milky Way
disk; therefore effective mechanisms must be at work to expel the
gas. \citet{freire2001} argue that only $0.5$\% of the spin-down
energy from the known millisecond pulsars in 47 Tuc would be required
to expel this large an amount of gas from the cluster. Other
mechanisms like novae and hot post-horizontal branch stars have also
been under discussion as responsible for the mass loss from globular
clusters. \citet{smith99} showed that fast stellar winds ejected by
main sequence stars can contribute to the gas loss. Therefore, at the
present time pollution seems unlikely to happen. The crucial question
then is whether during the era when intermediate mass stars evolved to
AGB stars material from their stellar winds could have been retained
in the globular cluster and have been re-accreted by other stars
including the low-mass main-sequence stars that are observed at the
present time. The mechanisms that are responsible for the ejection of
stellar winds out of globular clusters were likely already at work in
ancient times. Pulsars can form from stars with M=10 to 25 M\sunindex;
therefore many of them should have already formed by the time that 5
M\sunindex stars ejected their winds. Also main sequence star winds
should have been present in ancient times. Nonetheless,
\citet{thoul2002} show that the overwhelmingly largest
part of the gas from stellar winds is ejected within the first Gyr
after the birth of a GC, so that the accretion scenario may be
feasible in earlier times of cluster evolution.

\subsection{Other support for external pollution scenarios}

We have demonstrated that internal stellar evolution is unlikely to be
the dominant origin of the CN variations in 47 Tuc. While primordial
variations also have problems, the pollution scenario has its
attractions, although it remains unproven.

Recent work by other groups add strong arguments that external
enrichment processes can cause chemical inhomogeneities in GCs: from
high-resolution spectroscopy \citet{gratton01} found a Mg-Al
anti-correlation among turn-off stars in NGC\,6752. The core
temperatures of turn-off stars are too low to allow the proton capture
process to convert Mg to Al and cause this anti-correlation. The
authors conclude that internal mixing scenarios cannot account for the
chemical inhomogeneities in this globular cluster (which also exhibits
a bimodal CN distribution on the RGB \citealt{norris81}). They propose
external enrichment by AGB stars as the mechanism at work (see also
\citealt{ventura01}). In this regard, it is important to note that
\citet{briley96} had earlier shown that Na abundance inhomogeneities
exist among main sequence stars in 47 Tuc. Indeed, the relative
behavior of CN-Na-Mg-Al has a rich literature that we have not delved
into here. The review by \citet{kraft94} provides a good
introduction. Detailed abundance analysis of RGB stars in $\omega$ Cen
has revealed imprints of supernova Type Ia enrichment processes
\citep{pancino02}, but \citealt{cunha02} argue against this enrichment mode.
Furthermore, the mostly metal-poor $\omega$ Cen also contains a
metal-rich stellar population (e.g., \citealt{pancino00}) and has
consequently been proposed to be the core of a disrupted dwarf galaxy
(e.g., \citealt{freeman93}; \citealt{hughes00};
\citealt{gnedin02}). As such, it may be an atypical example of
self-enrichment processes in a globular cluster.

The question of the origin of the polluting gas in 47 Tuc can be
addressed by searching for the fingerprints of different enrichment
processes: cluster self-enrichment by massive stars versus accretion
of material ejected in winds from intermediate-mass stars. Do the
CN-strong main sequence stars have abundance enhancements in the
elements Na, Mg, and Al, which can be produced by proton-addition
reactions in intermediate-mass red giants, or by carbon burning in
even more massive stars? Do they show enhancements in the much heavier
r- and s-process elements such as Ba and Sr?  High resolution
spectroscopy might provide answers to these questions.

\acknowledgments

We thank the support staff of the ESO/VLT for expertly carrying out of
the FORS2/MXU observations in service observing mode. GHS gratefully
acknowledges the support of NSF grant AST 00-98453. We would like to
thank A. Burkert J. Gallagher for helpful discussions.




\clearpage

\clearpage

\clearpage

\clearpage

\clearpage

\clearpage

\clearpage

\clearpage

\clearpage
\begin{deluxetable}{rr@{\,\,}c@{\,\,}lr@{\,\,}c@{\,\,}lccccccc}
 \tablewidth{0pt}
 \tablecolumns{14}
 \tabletypesize{\small}
 \tablecaption{Data for Program stars \label{tab_stars}}
 \tablehead{
  \colhead{ID} & \multicolumn{3}{c}{$\alpha_{2000}$} &
  \multicolumn{3}{c}{$\delta_{2000}$} & \colhead{$V$} & \colhead{$(B-V)$} &
  \colhead{$\delta$\scnb} & \colhead{$\delta$\scnr} &
  \colhead{\sch} & \colhead{HK} &
  \colhead{Flag\tablenotemark{a} } }

 \startdata
   90 & 00 & 26 & 05.54 & -71 & 58 & 30.50 & 19.567 & 0.791 & 0.062 &  0.032 & 1.057 & 0.614 & 1 \\ 
  94 & 00 & 26 & 12.36 & -71 & 58 & 11.10 & 18.547 & 0.561 & 0.360 &  0.053 & 0.896 & 0.402 & 0 \\ 
 341 & 00 & 26 & 03.55 & -72 & 01 & 13.10 & 17.447 & 0.494 & 0.159 &  0.026 & 0.879 & 0.379 & 0 \\ 
 596 & 00 & 26 & 05.55 & -71 & 58 & 34.20 & 19.547 & 0.742 & 0.102 &  0.011 & 1.050 & 0.617 & 1 \\ 
 806 & 00 & 26 & 43.04 & -71 & 56 & 15.50 & 19.224 & 0.697 & 0.415 &  0.102 & 0.969 & 0.538 & 0 \\ 
 977 & 00 & 26 & 04.52 & -72 & 02 & 44.60 & 18.697 & 0.610 & 0.316 &  0.047 & 0.969 & 0.451 & 0 \\ 
1152 & 00 & 26 & 10.13 & -72 & 02 & 32.40 & 19.207 & 0.725 & 0.869 &  -0.035 & 0.893 & 0.697 & 2 \\ 
1173 & 00 & 26 & 20.17 & -71 & 52 & 12.50 & 17.662 & 0.514 & 0.030 &  -0.003 & 0.885 & 0.388 & 0 \\ 
1304 & 00 & 26 & 15.29 & -71 & 50 & 28.80 & 17.537 & 0.524 & 0.200 &  0.032 & 0.839 & 0.353 & 0 \\ 
1329 & 00 & 26 & 33.55 & -71 & 50 & 10.70 & 18.273 & 0.587 & 0.066 &  0.022 & 0.938 & 0.434 & 0 \\ 
1351 & 00 & 26 & 50.58 & -71 & 49 & 47.80 & 17.793 & 0.544 & 0.028 &  0.019 & 0.914 & 0.408 & 0 \\ 
1366 & 00 & 26 & 16.10 & -72 & 02 & 15.30 & 20.247 & 0.891 & 0.249 &  0.083 & 0.824 & 0.394 & 0 \\ 
1462 & 00 & 25 & 59.22 & -72 & 02 & 07.50 & 18.137 & 0.511 & 0.273 &  0.059 & 0.811 & 0.396 & 0 \\ 
1498 & 00 & 26 & 55.20 & -71 & 47 & 39.40 & 17.514 & 0.481 & 0.172 &  0.007 & 0.763 & 0.350 & 0 \\ 
1587 & 00 & 25 & 58.24 & -72 & 01 & 54.50 & 17.967 & 0.560 & 0.035 &  0.026 & 0.926 & 0.407 & 0 \\ 
1675 & 00 & 26 & 10.83 & -72 & 01 & 42.90 & 18.427 & 0.561 & -0.124 &  -0.021 & 1.059 & 0.817 & 2 \\ 
1833 & 00 & 26 & 04.95 & -72 & 01 & 24.10 & 19.707 & 0.758 & 0.056 &  0.022 & 1.055 & 0.597 & 2 \\ 
1890 & 00 & 26 & 03.53 & -72 & 01 & 16.10 & 17.727 & 0.527 & 0.420 &  -0.002 & 0.919 & 27.031 & 2 \\ 
1992 & 00 & 26 & 10.19 & -72 & 01 & 03.10 & 18.687 & 0.593 & 0.163 &  0.018 & 1.007 & 0.466 & 0 \\ 
2060 & 00 & 25 & 56.72 & -72 & 00 & 53.80 & 18.047 & 0.544 & 0.221 &  0.051 & 0.870 & 0.389 & 0 \\ 
2186 & 00 & 26 & 04.69 & -72 & 00 & 38.90 & 18.777 & 0.594 & 0.399 &  0.092 & 0.897 & 0.466 & 0 \\ 
2284 & 00 & 26 & 01.69 & -72 & 00 & 26.30 & 19.827 & 0.759 & 0.559 &  0.121 & 1.027 & 0.605 & 0 \\ 
2379 & 00 & 26 & 09.14 & -72 & 00 & 15.60 & 18.331 & 0.519 & 0.285 &  0.035 & 0.943 & 0.424 & 0 \\ 
2409 & 00 & 26 & 09.15 & -72 & 00 & 11.90 & 17.077 & 0.561 & 0.265 &  0.040 & 0.937 & 0.451 & 0 \\ 
2556 & 00 & 26 & 09.91 & -71 & 59 & 51.20 & 19.097 & 0.693 & 0.095 &  0.009 & 1.048 & 0.583 & 1 \\ 
2629 & 00 & 26 & 05.13 & -71 & 59 & 41.50 & 18.037 & 0.527 & 0.042 &  0.008 & 0.939 & 0.416 & 0 \\ 
2738 & 00 & 26 & 05.38 & -71 & 59 & 29.00 & 19.837 & 0.841 & 0.538 &  0.111 & 1.008 & 0.631 & 0 \\ 
2821 & 00 & 26 & 12.39 & -71 & 59 & 16.00 & 19.037 & 0.676 & 0.467 &  0.110 & 0.916 & 0.499 & 0 \\ 
2947 & 00 & 26 & 09.49 & -71 & 59 & 00.00 & 18.477 & 0.577 & 0.071 &  0.012 & 0.999 & 0.482 & 0 \\ 
3060 & 00 & 25 & 59.94 & -71 & 58 & 45.80 & 17.337 & 0.511 & 0.038 &  0.019 & 0.887 & 0.376 & 0 \\ 
3137 & 00 & 26 & 31.88 & -71 & 58 & 33.10 & 17.071 & 0.542 & 0.042 &  0.017 & 0.944 & 0.401 & 0 \\ 
3206 & 00 & 26 & 11.39 & -71 & 58 & 22.50 & 19.417 & 0.676 & 0.091 &  0.090 & 0.862 & 0.434 & 0 \\ 
3230 & 00 & 26 & 11.42 & -71 & 58 & 19.60 & 18.977 & 0.660 & 0.098 &  0.177 & -0.161 & 0.594 & 2 \\ 
3343 & 00 & 26 & 28.31 & -71 & 58 & 02.60 & 18.514 & 0.575 & 0.361 &  0.080 & 0.860 & 0.422 & 0 \\ 
3371 & 00 & 26 & 10.95 & -71 & 57 & 58.40 & 17.297 & 0.527 & 0.147 &  0.034 & 0.888 & 0.374 & 0 \\ 
3394 & 00 & 26 & 29.74 & -71 & 57 & 55.60 & 18.742 & 0.635 & 0.387 &  0.068 & 0.943 & 0.475 & 0 \\ 
3456 & 00 & 26 & 10.89 & -71 & 57 & 47.30 & 19.397 & 0.759 & 0.016 &  0.034 & 1.045 & 0.591 & 1 \\ 
3521 & 00 & 26 & 28.20 & -71 & 57 & 37.10 & 18.337 & 0.597 & 0.198 &  0.035 & 0.955 & 0.418 & 0 \\ 
3522 & 00 & 26 & 34.72 & -71 & 57 & 36.90 & 19.826 & 0.754 & 0.437 &  0.120 & 1.039 & 0.598 & 0 \\ 
3532 & 00 & 26 & 05.67 & -71 & 57 & 36.70 & 17.267 & 0.511 & 0.056 &  0.018 & 0.910 & 0.380 & 0 \\ 
3631 & 00 & 26 & 00.64 & -71 & 57 & 23.40 & 19.497 & 0.808 & 0.359 &  0.100 & 0.994 & 0.655 & 1 \\ 
3701 & 00 & 26 & 04.00 & -71 & 57 & 12.40 & 19.087 & 0.676 & 0.423 &  0.084 & 0.985 & 0.528 & 0 \\ 
3769 & 00 & 26 & 00.29 & -71 & 57 & 01.90 & 18.107 & 0.527 & 0.267 &  0.032 & 0.850 & 0.410 & 0 \\ 
3797 & 00 & 26 & 35.15 & -71 & 56 & 58.20 & 17.809 & 0.536 & 0.053 &  0.007 & 0.940 & 0.411 & 0 \\ 
3823 & 00 & 26 & 35.60 & -71 & 56 & 54.80 & 18.140 & 0.597 & 0.151 &  0.015 & 1.059 & 0.429 & 0 \\ 
3853 & 00 & 26 & 06.92 & -71 & 56 & 50.60 & 18.127 & 0.544 & 0.237 &  0.037 & 0.893 & 0.397 & 0 \\ 
3872 & 00 & 26 & 42.89 & -71 & 56 & 47.70 & 18.261 & 0.617 & 0.057 &  0.018 & 0.989 & 0.453 & 0 \\ 
3929 & 00 & 26 & 02.24 & -71 & 56 & 40.30 & 19.297 & 0.742 & 0.156 &  0.007 & 1.050 & 0.571 & 1 \\ 
3942 & 00 & 26 & 23.93 & -71 & 56 & 38.10 & 19.270 & 0.773 & 0.066 &  0.028 & 1.014 & 0.543 & 1 \\ 
4014 & 00 & 26 & 11.53 & -71 & 56 & 27.90 & 18.027 & 0.544 & 0.063 &  0.006 & 0.940 & 0.411 & 0 \\ 
4019 & 00 & 26 & 30.36 & -71 & 56 & 26.60 & 19.824 & 0.849 & 0.113 &  -0.001 & 1.019 & 0.595 & 2 \\ 
4023 & 00 & 26 & 13.09 & -71 & 56 & 26.50 & 19.482 & 0.772 & 0.465 &  0.109 & 0.983 & 0.548 & 1 \\ 
4115 & 00 & 26 & 11.27 & -71 & 56 & 12.60 & 17.877 & 0.527 & 0.058 &  0.016 & 0.919 & 0.420 & 0 \\ 
4188 & 00 & 26 & 12.11 & -71 & 56 & 03.10 & 19.687 & 0.758 & 0.150 &  0.040 & 1.054 & 0.723 & 1 \\ 
4298 & 00 & 26 & 54.92 & -71 & 55 & 44.30 & 18.422 & 0.608 & 0.287 &  0.068 & 0.928 & 0.444 & 0 \\ 
4344 & 00 & 26 & 33.67 & -71 & 55 & 37.70 & 17.607 & 0.527 & 0.150 &  0.035 & 0.859 & 0.383 & 0 \\ 
4421 & 00 & 26 & 34.06 & -71 & 55 & 26.00 & 18.631 & 0.602 & 0.376 &  0.066 & 0.934 & 0.461 & 0 \\ 
4452 & 00 & 26 & 17.37 & -71 & 55 & 21.80 & 18.605 & 0.547 & 0.017 &  -0.014 & 0.953 & 0.422 & 0 \\ 
4465 & 00 & 26 & 17.62 & -71 & 55 & 19.90 & 18.463 & 0.607 & 0.441 &  0.108 & 0.800 & 0.435 & 0 \\ 
4521 & 00 & 26 & 33.71 & -71 & 55 & 12.00 & 19.202 & 0.694 & 0.424 &  0.115 & 0.946 & 0.514 & 0 \\ 
4554 & 00 & 26 & 36.04 & -71 & 55 & 06.40 & 16.996 & 0.671 & 0.341 &  0.041 & 0.781 & 0.532 & 0 \\ 
4592 & 00 & 26 & 47.32 & -71 & 55 & 00.60 & 19.273 & 0.732 & 0.054 &  0.087 & 0.834 & 0.465 & 0 \\ 
4600 & 00 & 26 & 42.57 & -71 & 54 & 59.70 & 18.777 & 0.594 & 0.107 &  0.004 & 1.037 & 0.483 & 1 \\ 
4643 & 00 & 26 & 54.87 & -71 & 54 & 52.20 & 17.497 & 0.527 & 0.007 &  0.018 & 0.890 & 0.371 & 0 \\ 
4707 & 00 & 26 & 46.78 & -71 & 54 & 42.90 & 18.000 & 0.529 & 0.266 &  0.061 & 0.832 & 0.394 & 0 \\ 
4790 & 00 & 26 & 56.83 & -71 & 54 & 27.50 & 18.437 & 0.577 & 0.237 &  0.028 & 0.947 & 0.438 & 0 \\ 
4877 & 00 & 26 & 46.37 & -71 & 54 & 14.00 & 18.457 & 0.659 & 0.297 &  0.052 & 0.956 & 0.452 & 0 \\ 
4936 & 00 & 26 & 53.02 & -71 & 54 & 03.20 & 17.937 & 0.544 & 0.248 &  0.041 & 0.866 & 0.401 & 0 \\ 
5049 & 00 & 26 & 50.13 & -71 & 53 & 43.50 & 17.007 & 0.626 & 0.199 &  0.023 & 0.973 & 0.432 & 0 \\ 
5146 & 00 & 26 & 54.38 & -71 & 53 & 26.80 & 19.949 & 0.851 & 0.357 &  0.044 & 1.006 & 0.519 & 0 \\ 
5154 & 00 & 26 & 48.22 & -71 & 53 & 26.00 & 18.377 & 0.561 & 0.040 &  -0.004 & 0.973 & 0.413 & 0 \\ 
5188 & 00 & 27 & 04.02 & -71 & 53 & 18.70 & 17.357 & 0.527 & 0.181 &  0.054 & 0.795 & 0.359 & 0 \\ 
5205 & 00 & 26 & 19.88 & -71 & 53 & 16.60 & 17.222 & 0.513 & 0.231 &  0.058 & 0.775 & 0.371 & 0 \\ 
5214 & 00 & 26 & 51.50 & -71 & 53 & 13.20 & 19.731 & 0.688 & 0.388 &  4.746 & 0.799 & 0.401 & 0 \\ 
5219 & 00 & 26 & 20.25 & -71 & 53 & 13.50 & 18.528 & 0.589 & 0.026 &  0.005 & 0.971 & 0.468 & 0 \\ 
5231 & 00 & 26 & 51.88 & -71 & 53 & 10.60 & 18.737 & 0.626 & 0.445 &  0.113 & 1.024 & 0.552 & 1 \\ 
5235 & 00 & 26 & 56.51 & -71 & 53 & 09.80 & 19.847 & 0.841 & 0.172 &  -0.007 & 1.041 & 0.628 & 2 \\ 
5237 & 00 & 26 & 31.91 & -71 & 53 & 09.50 & 19.255 & 0.744 & 0.046 &  0.007 & 1.040 & 0.593 & 0 \\ 
5253 & 00 & 26 & 36.09 & -71 & 53 & 07.60 & 18.879 & 0.658 & 0.475 &  0.069 & 0.932 & 0.488 & 0 \\ 
5261 & 00 & 26 & 21.88 & -71 & 53 & 06.80 & 20.255 & 0.864 & 0.161 &  0.041 & 1.066 & 0.569 & 0 \\ 
5296 & 00 & 27 & 15.20 & -71 & 52 & 58.70 & 19.077 & 0.676 & 0.047 &  0.082 & 0.850 & 0.448 & 0 \\ 
5299 & 00 & 26 & 23.93 & -71 & 53 & 00.30 & 19.153 & 0.668 & 0.113 &  0.011 & 1.046 & 0.591 & 1 \\ 
5342 & 00 & 27 & 03.51 & -71 & 52 & 51.40 & 17.477 & 0.511 & 0.178 &  0.029 & 0.813 & 0.369 & 0 \\ 
5381 & 00 & 26 & 54.82 & -71 & 52 & 42.70 & 19.047 & 0.676 & 0.419 &  0.048 & 0.936 & 0.471 & 0 \\ 
5402 & 00 & 26 & 35.43 & -71 & 52 & 38.60 & 18.377 & 0.610 & 0.053 &  0.021 & 0.977 & 0.464 & 0 \\ 
5538 & 00 & 27 & 02.38 & -71 & 52 & 12.80 & 18.317 & 0.577 & 0.186 &  0.022 & 0.959 & 0.428 & 1 \\ 
5545 & 00 & 26 & 35.31 & -71 & 52 & 11.50 & 18.636 & 0.638 & 0.063 &  0.006 & 1.004 & 0.482 & 0 \\ 
5547 & 00 & 26 & 52.68 & -71 & 52 & 10.60 & 19.397 & 0.775 & 0.643 &  0.136 & 0.954 & 0.536 & 0 \\ 
5587 & 00 & 26 & 44.72 & -71 & 52 & 01.60 & 17.701 & 0.549 & 0.029 &  0.013 & 0.926 & 0.409 & 0 \\ 
5653 & 00 & 26 & 30.71 & -71 & 51 & 47.60 & 18.046 & 0.562 & 0.190 &  0.038 & 0.871 & 0.397 & 0 \\ 
5665 & 00 & 26 & 33.74 & -71 & 51 & 43.60 & 17.154 & 0.544 & 0.229 &  0.061 & 0.790 & 0.394 & 0 \\ 
5759 & 00 & 26 & 14.33 & -71 & 51 & 26.10 & 17.318 & 0.514 & 0.178 &  0.036 & 0.836 & 0.360 & 0 \\ 
5809 & 00 & 26 & 48.71 & -71 & 51 & 16.90 & 19.243 & 0.724 & 0.050 &  0.006 & 1.035 & 0.583 & 0 \\ 
5836 & 00 & 26 & 35.80 & -71 & 51 & 11.60 & 18.757 & 0.658 & 0.311 &  0.036 & 0.991 & 0.506 & 0 \\ 
5942 & 00 & 26 & 52.82 & -71 & 50 & 48.50 & 19.143 & 0.740 & 0.525 &  0.099 & 0.980 & 0.550 & 0 \\ 
5950 & 00 & 26 & 33.32 & -71 & 50 & 47.50 & 19.096 & 0.716 & 0.420 &  0.530 & -0.000 & -0.000 & 2 \\ 
6043 & 00 & 26 & 48.02 & -71 & 50 & 24.60 & 17.810 & 0.527 & 0.229 &  0.064 & 0.803 & 0.385 & 0 \\ 
6047 & 00 & 26 & 51.54 & -71 & 50 & 23.80 & 17.118 & 0.562 & 0.036 &  0.029 & 0.917 & 0.404 & 0 \\ 
6220 & 00 & 26 & 39.76 & -71 & 49 & 42.70 & 18.537 & 0.691 & 0.381 &  0.103 & 0.856 & 0.459 & 0 \\ 
6248 & 00 & 26 & 52.04 & -71 & 49 & 35.70 & 18.919 & 0.734 & 0.094 &  0.009 & 1.014 & 0.543 & 0 \\ 
6264 & 00 & 26 & 40.87 & -71 & 49 & 33.80 & 19.871 & 0.757 & 0.419 &  0.112 & 0.978 & 0.671 & 1 \\ 
6335 & 00 & 27 & 02.56 & -71 & 49 & 16.80 & 19.207 & 0.699 & 0.075 &  0.093 & 0.812 & 0.482 & 0 \\ 
6380 & 00 & 26 & 50.93 & -71 & 49 & 09.30 & 19.685 & 0.735 & 0.513 &  0.111 & 0.976 & 0.614 & 0 \\ 
6452 & 00 & 26 & 48.80 & -71 & 48 & 50.70 & 18.507 & 0.626 & 0.355 &  0.037 & 0.922 & 0.447 & 0 \\ 
6455 & 00 & 26 & 58.61 & -71 & 48 & 49.70 & 17.901 & 0.574 & 0.190 &  0.030 & 0.888 & 0.396 & 0 \\ 
6537 & 00 & 26 & 32.22 & -71 & 48 & 29.00 & 17.198 & 0.549 & 0.106 &  0.042 & 0.887 & 0.391 & 0 \\ 
6591 & 00 & 26 & 55.56 & -71 & 48 & 14.60 & 17.444 & 0.534 & 0.186 &  0.034 & 0.823 & 0.366 & 0 \\ 
6622 & 00 & 27 & 09.57 & -71 & 48 & 04.80 & 21.893 & 0.219 & 0.071 &  -0.039 & 0.729 & 0.272 & 2 \\ 
6641 & 00 & 27 & 13.08 & -71 & 48 & 00.70 & 19.275 & 0.739 & 0.522 &  0.106 & 0.931 & 0.583 & 0 \\ 
6648 & 00 & 26 & 45.61 & -71 & 47 & 59.50 & 19.091 & 0.684 & 0.505 &  0.088 & 0.980 & 0.507 & 0 \\ 
6723 & 00 & 26 & 48.58 & -71 & 47 & 39.90 & 19.086 & 0.693 & 0.543 &  0.056 & 0.973 & 0.538 & 0 \\ 

 \enddata
 \tablecomments{\tablenotemark{a} Qualityflag: 0=good quality spectra; 
  1=some ambiguities; 2=poor quality spectra - unreliable.}
\end{deluxetable}

\end{document}